\documentclass[12pt]{spieman}  

\usepackage{amsmath,amsfonts,amssymb}
\usepackage{graphicx}
\usepackage{setspace}
\usepackage{tocloft}

\usepackage[utf8]{inputenc}
\usepackage[T1]{fontenc}
\usepackage{esdiff} 
\usepackage{wrapfig}
\usepackage{mathrsfs}

\title{Progress toward optimizing energy and arrival-time resolution with a transition-edge sensor from simulations of X-ray-photon events}

\author[a,*]{Paul Ripoche}
\author[a]{Jeremy Heyl}
\affil[a]{University of British Columbia, Department of Physics and Astronomy, 6224 Agricultural Road, Vancouver, British Columbia, Canada, V6T 1Z4}

\cftpagenumbersoff{figure}
\cftpagenumbersoff{table} 
\begin{document} 
\maketitle

\begin{abstract}
Superconducting transition-edge sensors (TESs) carried by X-ray telescopes are powerful tools for the study of neutron stars and black holes. Several methods, such as optimal filtering or principal component analysis, have already been developed to analyse X-ray data from these sensors. However, these techniques may be hard to implement in space.
Our goal is to develop a lower-computational-cost technique that optimizes energy and time resolution when X-ray photons are detected by a TES. 
TESs exhibit a non-linear response with photon energy. Therefore, at low energies we focus on the current-pulse height whereas at high energies we consider the current-pulse width, to retrieve energy and arrival time of X-ray photons.
For energies between $0.1$ keV and $30$ keV and with a sampling rate of $195$ kHz, we obtain an energy resolution (full width at half the maximum) between $1.32$ eV and $2.98$ eV. We also get an arrival-time resolution (full duration at half the maximum) between $163$ ns and $3.85$ ns.
To improve the accuracy of these results it will be essential to get a thorough description of non-stationary noise in a TES, and to develop a robust on-board identification method of pile-up events. 
\end{abstract}


{\noindent \footnotesize\textbf{*}Paul Ripoche,  \linkable{pripoche@phas.ubc.ca} }

\begin{spacing}{2}   

\section{Introduction}

X-ray telescopes enable us to study fascinating compact objects, such as neutron stars and black holes. In the future, several missions, e.g. Athena \cite{2018SPIE10699E..1GB}, Colibrì \cite{2019BAAS...51g.175H} or Lynx \cite{2018SPIE10699E..0NG}, are planned to carry arrays of superconducting transition-edge sensors (TESs). Consequently, optimizing energy resolution and timing for photons detected with such sensors is a crucial goal in X-ray astronomy \cite{2015SuScT..28h4003U}.

Recently, several techniques to analyze X-ray data from transition-edge sensors have been developed, such as principal component analysis \cite{2020JLTP..199..745F, 2016JLTP..184..382B, 2016SPIE.9905E..5VD}, optimal filtering of the resistance signal \cite{2015ApPhL.107v3503L} and optimal filtering analysis \cite{2019JATIS...5b1008S, 2016JLTP..184..374F, 2014AIPA....4k7106S, 2005NIMPA.555..255W, 2004NIMPA.520..592W, 2004NIMPA.520..555F, 2002AIPC..605..339F, 2000NIMPA.444..453F}. These techniques achieve an energy resolution between $0.7$ eV and $3.4$ eV full width at half maximum (FWHM), for low energies (below $6$ keV). However, these techniques may be difficult to implement easily on an X-ray space telescope. Indeed, aboard an X-ray telescope one is limited by processor power consumption and hardware reliability in a radiation environment; the latter limits the memory and processing speed because of the limited range of components that are rated for use in space. Furthermore, one is limited by data bandwidth to the ground, so nearly all of the pulses must be processed on board.

Therefore, we propose a method easy to implement on an X-ray telescope, in order to get the energy and the arrival time of photons detected by a TES. We aim to use this technique to predict the behavior of the new generation of non-focusing X-ray telescopes. Indeed, such X-ray telescopes will be studying variable X-ray emissions coming from neutron stars and black holes, and will probe the region very close to them, where the dynamical timescales of those region are of the order of microseconds. Since accreting neutron stars and black holes shine bright in the $0.5-10$ keV range, we developed a technique that optimizes time and energy resolution over the widest range of photon energies (here, from $0.1$ keV to $30$ keV). 
 
This paper first describes our detector model and how events were simulated. We next address the issue of noise in TESs and present our method that enables to optimize energy resolution and timing.  Finally, we discuss the outcomes.

\section{Transition-edge sensors to detect X-rays}
\subsection{Detector model}
\label{sec:detectModel}
A transition-edge sensor is made of a superconducting metal film functioning near its transition temperature $T_c$ (typically $0.1$ K). While electrons move freely in a superconducting metal, they encounter some significant resistance when the metal switches to its normal phase. The transition from superconductor to normal metal occurs within about a narrow 1~mK change in the temperature, but results in a large change in resistance. Thus after an X-ray photon deposits energy in the sensor, the superconductor heats up, the resistance increases, the electric current drops and an X-ray photon is detected.

\begin{wrapfigure}{R}{0.4\columnwidth}
    \begin{center}
        \includegraphics[width=0.4\columnwidth]{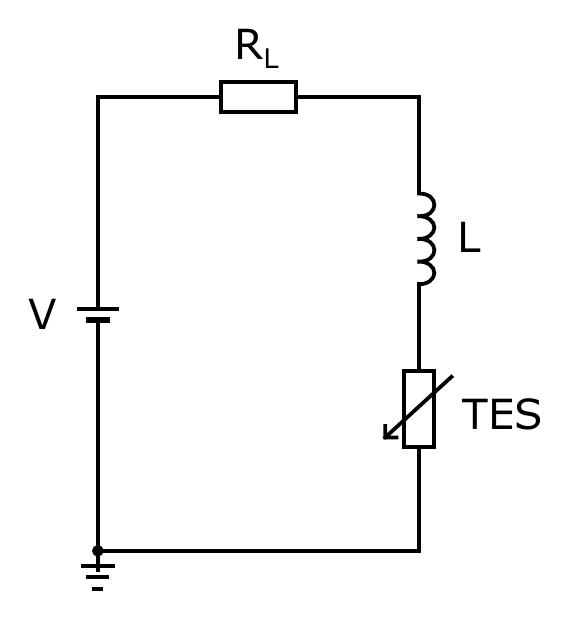}
    \end{center}
    \caption{Thevenin-equivalent representation of a TES input circuit.}
    \label{fig:electrocircuit}
\end{wrapfigure}

Indeed, when an X-ray photon hits a TES, the provided energy provokes the needed rise in temperature for the transition to happen; therefore measuring how much the current diminishes  and how long the sensor takes to recover enables us to determine the energy of that photon. Finally, after the absorption of a photon, a TES needs to be cooled down to its initial temperature by using a cooling bath, at temperature $T_\text{bath}$, in order to be able to detect the next photon. The current in a TES is measured with a SQUID readout circuit (see Fig. 2 in \cite{Irwin2005} for more details).
In this work, we assume that the response of a TES is described by two coupled differential equations governing the electrical and thermal circuits \cite{Irwin2005}. In Fig.~\ref{fig:electrocircuit}, we provide a Thevenin-equivalent representation of a TES input circuit. For both equations, we first ignore noise terms and the signal power. The electrical current intensity $I$ is governed by the electrical differential equation:
\begin{equation}
    L\diff{I}{t} = V - IR(T,I) - IR_L,
    \label{eq:Idyn}
\end{equation}
where $R(T, I)$ is the electrical resistance of a TES (a function of both temperature and current), $L$ is the SQUID input inductance, $V$ is the Thevening-equivalent bias voltage, and $R_L$ is a load resistor. The temperature $T$ is governed by the thermal differential equation:
\begin{equation}
    C\diff{T}{t} = I^{2}R(T,I) - k(T^{n} - T_\text{bath}^{n}),
    \label{eq:Tdyn}
\end{equation}
where $n$ is the thermal conductance exponent, $C$ is the heat capacity of both the TES and any absorber, and $k = G/(nT_0^{n-1})$ -- where $G \equiv \text{d}[k(T^{n} - T_\text{bath}^{n})]/\text{d}T |_0$ is the thermal conductance to the heat
bath. While empirically mapping out $R(T, I)$ for superconducting transitions is an undergoing research \cite{2020JPhCS1590a2032Z,2018JLTP..193..321Z,2017JAP...121g4503Z}; we use a model \cite{2014AIPA....4k7106S,2007PhDT.........3B}, motivated by Ginzburg-Landau theory, that is well adapted to simulations:
\begin{equation}
    R(T,I) = \frac{R_N}{2} \left \{ 1 + \tanh\left[\frac{T-T_C + (I/A)^{2/3}}{2\ln(2)T_W}\right] \right \},
    \label{eq:Rdetect}
\end{equation}
where $R_N$ is the normal resistance of a TES, $T_W$ is the $10$-$90$\% transition width, and $A$ is the strength of the suppression of $T_c$ by non-zero current density in the superconducting metal film.

The two differential equations contain nonlinear terms in both temperature and current, inherent to the bias input circuit of a TES. Nonlinearities also occur during the transition, hence a nonlinear TES resistance. 

We assume that the TES parameters in Eq.~(\ref{eq:Idyn}), Eq.~(\ref{eq:Tdyn}) and Eq.~(\ref{eq:Rdetect}) are the same as for detectors being developed at NIST (National Institute of Standards and Technology) for the HOLMES experiment. Those parameters are summarized in \cite{2016JLTP..184..263A} and are reported in Table~\ref{tab:physparam}. The steady-state values (at quiescence) of resistance, temperature, and current are, respectively: $R_0$ , $T_0$ , $I_0$. The unitless logarithmic temperature sensitivity and current sensivity, respectively $\alpha_I$ and $\beta_I$ are defined as:
\begin{equation}
    \begin{aligned}
        \alpha_I \equiv \frac{\partial \log R}{\partial \log T} \bigg|_{I_0}, \\
        \beta_I \equiv \frac{\partial \log R}{\partial \log I} \bigg|_{T_0}.
    \end{aligned}
\end{equation}

\begin{table}[ht]
    \caption{\label{tab:physparam} TES parameters used for detectors being developed at NIST for the HOLMES experiment \cite{2016JLTP..184..263A}.}
    \begin{center}
    \begin{tabular}{|l|l|}
        \hline
        \multicolumn{2}{|c|}{\textbf{TES parameters}} \\
        \hline
        \multicolumn{1}{|c|}{Assumed} &  \multicolumn{1}{|c|}{Derived} \\
        \hline
        $n=3.25$ &  $V=146.9$ nV \\
        $k=23.3$ nW.K$^{-n}$ & $T_W = 0.565$ mK \\
        $T_c=0.1$ K & $A = 1.133$ A.K$^{-3/2}$ \\
        $C=0.5$ pJ.K$^{-1}$ & $T_0=0.0980$ K \\
        $R_L=0.3$ m$\Omega$ & $I_0 = 63.85$ $\mu$A \\
        $T_\text{bath}=0.07$ K & $G = 406.8$ pW.K$^{-1}$ \\
        $R_N=10$ m$\Omega$ & \\
        $R_0=2$ m$\Omega$ & \\
        $L \in \{12, 24, 48\}$ nH & \\
        $\alpha_I = 200$ & \\
        $\beta_I = 2$ & \\
        \hline
    \end{tabular}
    \end{center}
\end{table}

\subsection{Motivations}
\label{sec:simulEvents}

In order to develop a method that gives the energy and the arrival time of incoming photons, while optimizing the resolution for those two parameters, we simulated single events.

\begin{figure}[ht]
\includegraphics[width=\columnwidth]{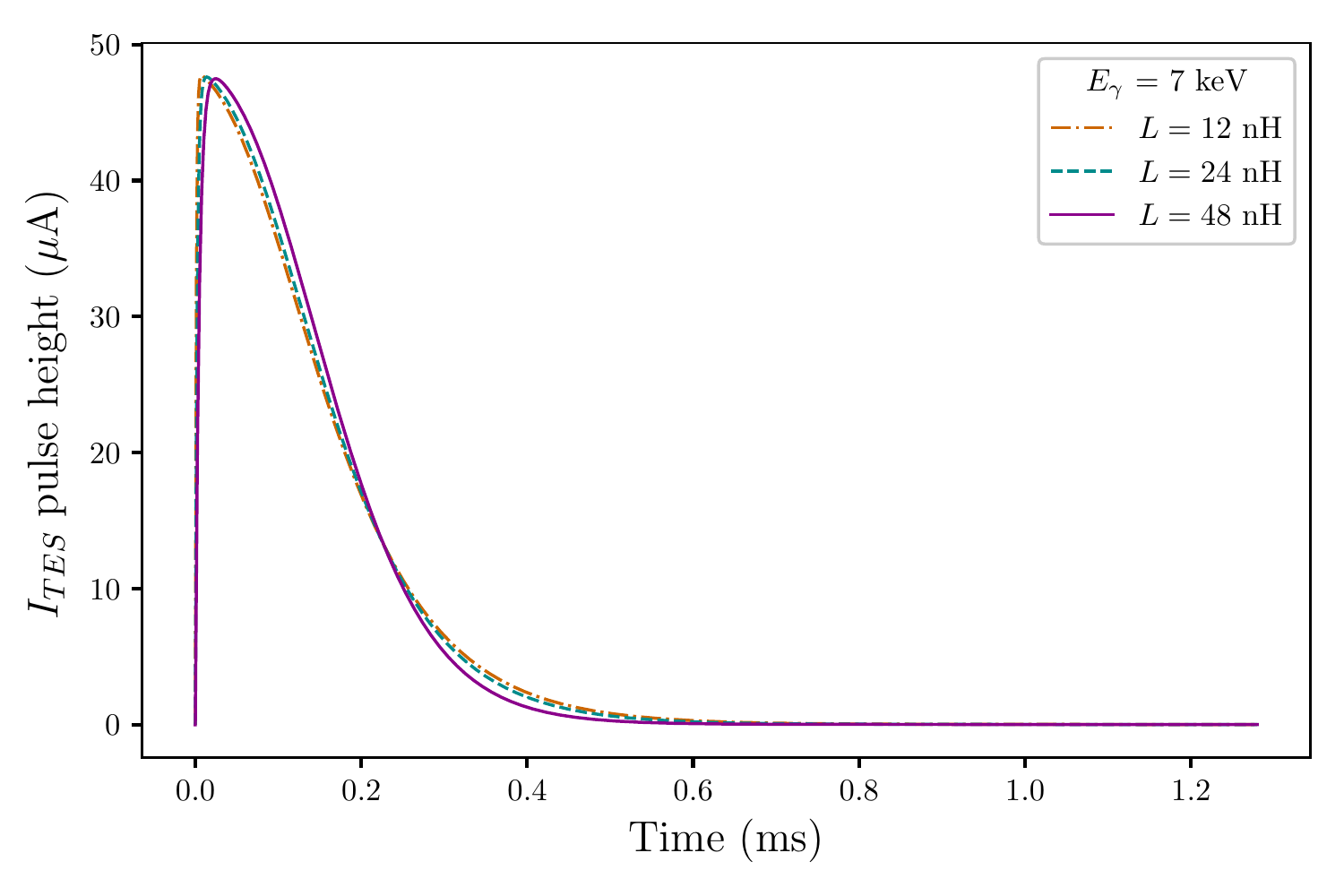}
\caption{Current pulse in a TES for different values of inductance. The incoming photon has an energy of $7$ keV.}
\label{fig:current_pulse_7keV}
\end{figure}

To simulate each photon arrival, we solved Eq.~(\ref{eq:Idyn}) and Eq.~(\ref{eq:Tdyn}) letting the initial temperature and current of the TES be:
\begin{equation}
    T_{0,\gamma}= T_0 + \frac{E_\gamma}{C},
\end{equation}
\begin{equation}
    I_{0,\gamma} \equiv I_0,
\end{equation}
where $E_\gamma$ is the energy of each simulated incoming photon. The current pulse height evolution is shown in Fig.~\ref{fig:current_pulse_7keV} for a $7$~keV photon.

One can notice from Fig.~\ref{fig:current_pulse_7keV}, that the rising time increases with the inductance. So, throughout this work, except otherwise mentioned, we use $L=48$ nH, because it allows lower sampling rates in our methods.

\begin{figure}[ht]
\includegraphics[width=\columnwidth]{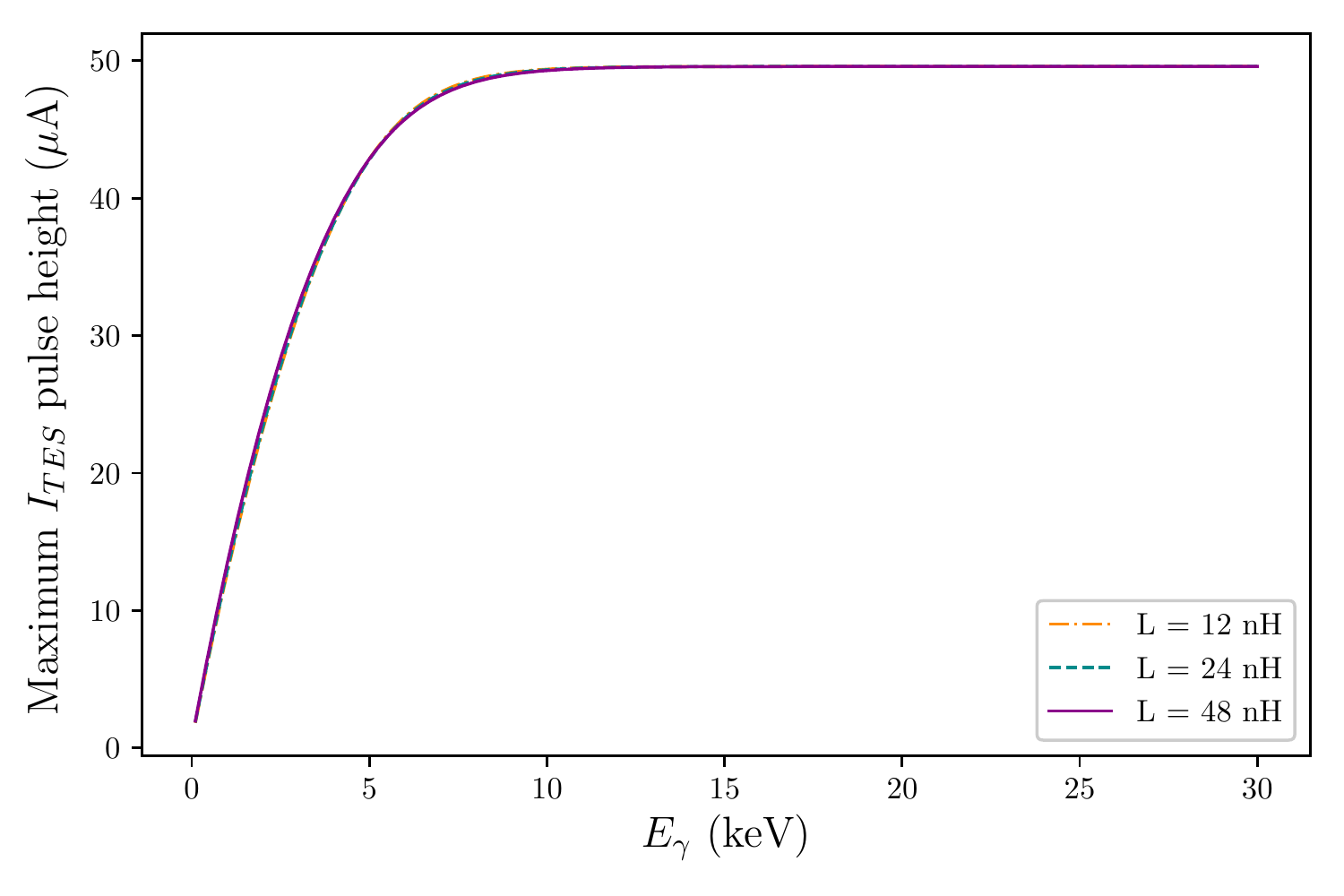}
\caption{Maximum current-pulse height in a TES as a function of the energy of the incoming photon. We find the behaviour to be similar for different inductance values.}
\label{fig:max_pulse_height}
\end{figure}

\begin{figure}[ht]
\includegraphics[width=\columnwidth]{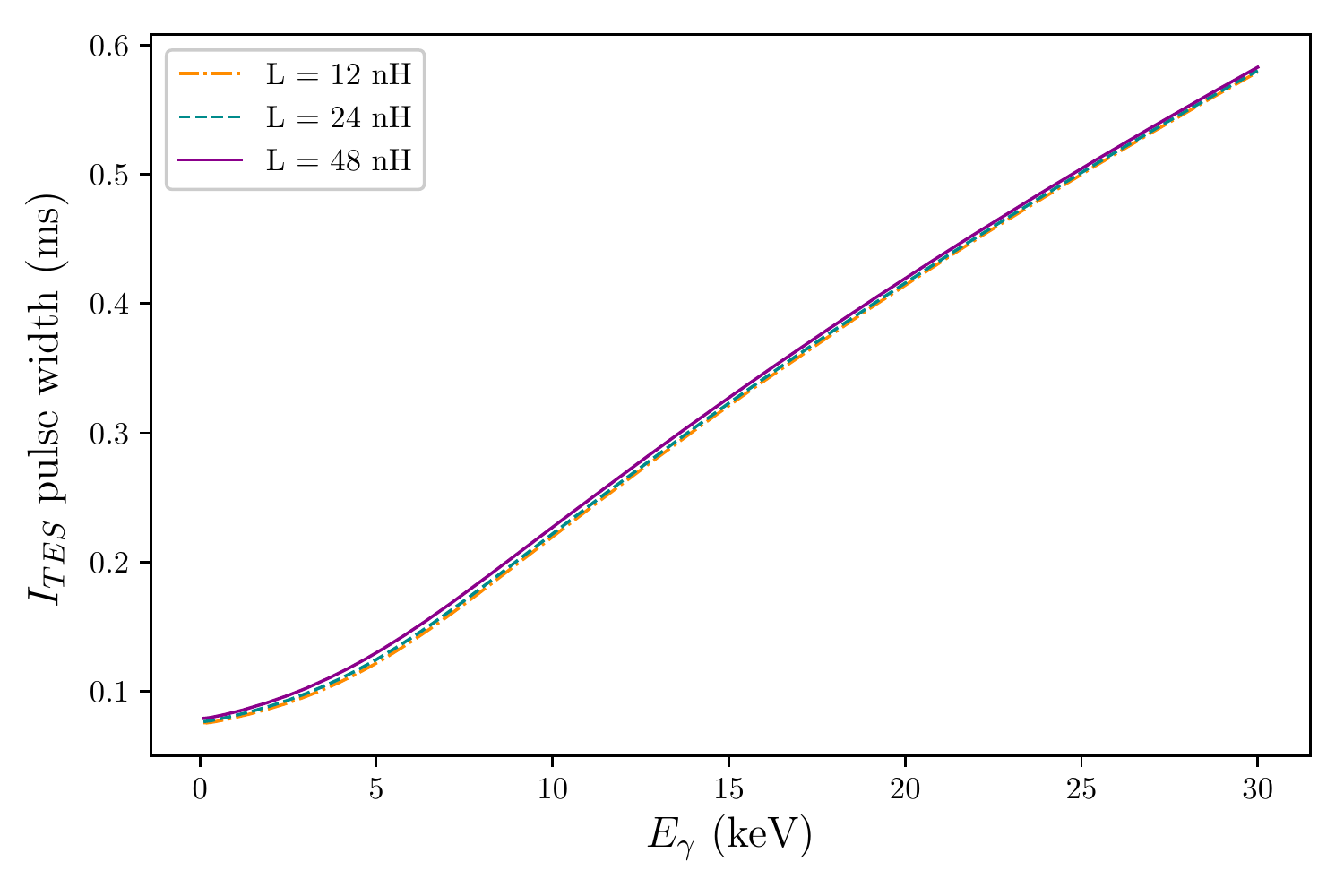}
\caption{Current-pulse width in a TES as a function of the energy of the incoming photon. We find the behaviour to be similar for different inductance values.}
\label{fig:pulse_width}
\end{figure}

We then generated single events at energies in the $0.1-30$~keV range, for the three different values of inductance (see Table~\ref{tab:physparam}). We observe that the maximum current-pulse height in the TES increases with the energy of the incoming photon, until it clearly saturates for $E_\gamma \geq 10$ keV; this is illustrated in Fig.~\ref{fig:max_pulse_height}. Once the pulse flattens at high energies, the fall time increases significantly with the energy of the incoming photon. So we also measured the width of the current pulse and plotted it against the energy of the incoming photon, see Fig.~\ref{fig:pulse_width}. We then infer from those previous two observations that the current-pulse height has a linear behavior for $E_\gamma \leq 4$ keV, whereas the width of the current signal becomes linear for $E_\gamma \geq 4$ keV. In other words, there exists a switching-point energy that determines which property of the current pulse in a TES should be used to measure the energy of the incoming photon and its arrival time. 

These observations from our simulations have been well studied in theory. Indeed, in a TES, when the energy of the incoming photon exceeds the saturation energy of the detector, $E_{sat} = C T_0/\alpha_I$, the top part of the current pulse flattens \cite{Irwin2005}. Consequently, the energy resolution is degraded for $E_{\gamma} \geq E_{sat}$. In our case, we calculate $E_{sat} = 1.53$ keV. However the energy of the incoming photon can still be estimated at higher energies, since the current pulse saturates for longer period of times (the width of the current pulse increases with the energy). This is valid as long as the pulse fall time is much smaller than the natural thermal time constant $\tau \equiv C/G$. We find $\tau = 1.23$ ms, for our TES. Finally, when the fall time reaches $\tau$ the TES current response becomes insensitive to the photon energy.

The magnitude of X-ray pulses as a function of energy has already been used to retrieve the energy of an incoming photon \cite{2013ITAS...2300705B}. As mentioned before, several techniques, such as optimal filtering or principal component analysis, already achieve energy resolutions below $3.5$ eV, at low energies. However, at higher energies, the resolution degrades from that obtained at low energies, because TESs have a very non-linear response and because the noise is generally non-stationary (see section~\ref{sec:detectModel}). Optimal analysis for large pulses can be used, but it becomes a much more laborious process than at low energies \cite{2004NIMPA.520..555F} (different filter function templates for different energies). Consequently, to optimize time and energy resolution for X-ray photons, we propose to treat high-energy and low-energy photons with two different techniques. In general, the switching-point energy depends on the properties of the detector and corresponds to the energy of the photon that begins to saturate the superconductor to normal transition of the TES.

\section{Noise in a transition-edge sensor}
\label{sec:noise}
\subsection{Different noise sources}
As in all physical dissipated systems, the current and temperature of a TES are affected by thermodynamic fluctuations. These thermodynamic noise sources are thermal fluctuation noise (also known as phonon noise), and Johnson–Nyquist noise \cite{McCammon2005, Irwin2005}. Thermal fluctuation (TF) noise is the statistical fluctuations that arise from energy exchange between the detector and the heat sink. Johnson–Nyquist (JN) noise is produced by the thermal agitation of electrons in a TES, in other words, it is the electronic noise at equilibrium. The energy resolution of a TES is fundamentally limited by these two noise sources. From these fundamental limits, the energy resolution can be degraded by additional noise sources. Such sources \cite{Irwin2005} can be quantum fluctuations (usually negligible in a TES), fluctuations in the superconducting order parameter, flux motion, and internal thermodynamic fluctuations (due to poorly-coupled heat capacity for example). The latter is usually negligible compared to all other noise sources by use of low resistance TESs; we also neglect the other additional noise sources, in this study.

Finally, measurements in a typical TES show the presence of an excess noise compared to the noise predictions. This excess noise has the same frequency dependence as the TES Johnson–Nyquist noise \cite{2013JAP...114g4513S,2008JLTP..151..119K,2004ApPhL..84.4206U}, and can be separated from internal thermodynamic fluctuation noise \cite{2019JAP...125p4503W}. Because the nature of this excess noise is not yet fully understood, its magnitude is often assumed to be an additional term in the TES JN noise. In other words, one can define a factor of $(1+m^2)$ multiplying the TES JN noise, where $m$ is a dimensionless scale factor that quantifies the magnitude of the excess noise \cite{2019JAP...125p4503W,2013JAP...114g4513S,2004ApPhL..84.4206U}. Furthermore, in an ideal linear TES, Johnson–Nyquist noise is also produced by the SQUID amplifier and the load resistor ($R_L$) \cite{Irwin2005}.

\subsection{Adding noise to our simulations}
The noise sources considered for our study are Johnson–Nyquist noise in the TES and in the load resistor, thermal fluctuation noise, and the noise of the SQUID readout circuit. SQUID amplifiers have both current and voltage noise which are strongly correlated. However, correlated voltage noise terms can be neglected because the impedance of the TES is usually higher than the noise impedance of the SQUID \cite{Irwin2005}. 

Since we simulate TES current signals, it is more appropriate for our analysis to use current noises. Several assumptions are needed to be able to do so. Indeed, the TES resistance has both temperature-dependent and current-dependent nonlinearity that can change the random noise voltage across the resistor. Specifically, in circuits with current-dependent nonlinearity the Thevenin theorem does not apply, so the TES JN noise which has a series voltage source does not have an equivalent parallel current noise source \cite{Irwin2005}. Moreover, a nonlinear resistance has non-Gaussian noise \cite{Irwin2005}. Consequently, for our noise calculations we assume the nonlinear equilibrium ansatz (NLEA) \cite{Irwin2005} which allows the resistor to have current-dependent nonlinearity ($\beta_I \neq 0$), but the noise is still calculated assuming a system near equilibrium. 

Furthermore, for large pulses, the noise is generally non-stationary, especially at the top of the pulse because the TES has a higher resistance. However a full description of non-stationary noise is beyond the scope of this paper, and we instead calculate current noise signals in the small signal limit, assuming a quadratic nonlinear TES resistor \cite{Irwin2005}.

In order to add noise to our TES current simulations, we start in the frequency domain, considering power spectral densities, which we then convert into real (as opposed to complex) time-domain current noise signals. Finally, since we assume that all the noises are uncorrelated, we can add each noise signal to the TES current at every time step of the simulation. Another approach could be to directly add randomly generated white noise (with the appropriate variance for each noise source) in the two coupled differential equations governing the electrical and thermal circuits \cite{2016SPIE.9905E..64W}.
We use the power spectral densities of the current noises described in Table 1 of Irwin (2005). The power spectral density of the current noise due to JN noise voltages in the TES is:
\begin{equation}
    S_\text{TES}(\omega)= 4 k_B T_0 I_0^2 R_0 \frac{(1+2\beta_I)}{\mathscr{L}_I^2} (1 + \omega^2 \tau^2)|s_I(\omega)|^2,
\label{eq:PSD_JN_TES}
\end{equation}
where $k_B$ is the Boltzmann constant, $\mathscr{L}_I$ is the low-frequency loop gain under constant current, $\tau \equiv C/G$ is the natural thermal time constant, $\omega = 2 \pi f$ with $f$ the frequency, and $s_I(\omega)$ is the power-to-current responsivity. In our simulations, we multiply $S_\text{TES}$ by $(1+m^2)$ to account for the excess noise. Following measurements for the HOLMES experiment \cite{2019JInst..14P0035B}, we take $m=1.5$, in this analysis.
The power spectral density of the current noise due to JN noise voltages in the load resistor is:
\begin{equation}
    S_\text{load}(\omega)= 4 k_B T_\text{bath} I_0^2 R_L \frac{(\mathscr{L}_I-1)^2}{\mathscr{L}_I^2} (1 + \omega^2 \tau_I^2)|s_I(\omega)|^2,
\label{eq:PSD_JN_load}
\end{equation}
where we assume the load resistor to be at the same temperature as the cooling bath, and $\tau_I$ is the constant-current time constant.
The power spectral density of the current-noise fluctuations due to TF noise is:
\begin{equation}
    S_\text{TF}(\omega)= 4 k_B T_0^2 G F(T_0, T_\text{bath})|s_I(\omega)|^2,
\label{eq:PSD_TF}
\end{equation}
where the form of the unitless function $F(T_0, T_\text{bath})$ depends on the thermal conductance exponent and on whether phonon reflection from the boundaries is radiative or diffuse. It is defined as \cite{McCammon2005} $F(T_0, T_\text{bath}) = F_\text{LINK}(T_\text{bath}, n) (T_\text{bath}/T_0)^{n+1}$. We assume the radiative limit in this work and $F_\text{LINK}$
 is given by \cite{McCammon2005,1959JOSA...49...66B}: 
 \begin{equation}
    F_\text{LINK} = \frac{(T_0/T_\text{bath})^{n+2} + 1}{2},
\label{eq:PSD_SQUID}
\end{equation}
which gives, $F_\text{LINK} = 0.82$; in the diffusive limit \cite{McCammon2005,1982ApOpt..21.1125M}, we get $F_\text{LINK} = 0.79$, so the choice of limit should not impact our  noise analysis. 
The power spectral density of the current noise due to noise from SQUID amplifiers, for the HOLMES experiment, is taken to be \cite{2019JInst..14P0035B, 2019EPJC...79..304A}:
\begin{equation}
    S_\text{SQUID} = 4 \times 10^{-22}\mathrm{A}^2.\mathrm{Hz}^{-1}.
\label{eq:PSD_JN_squid}
\end{equation}
 We invite the reader to refer to Table 1 from Irwin (2005) for the expression of the parameters not defined in the above equations.

\begin{table}[H]
    \caption{\label{tab:simuparam} Sampling parameters used for simulations of events in a TES.}
    \begin{center}
    \begin{tabular}{|c|c|c|}
        \hline
        \multicolumn{3}{|c|}{\textbf{Sampling parameters}} \\
        \hline
        \multicolumn{3}{|c|}{Record length: $T_s=1.28$ ms; frequency resolution: $\Delta f = 781.25$ Hz} \\
        \hline
        \multicolumn{1}{|c|}{Number of samples:} &  \multicolumn{1}{|c|}{Sample interval:} &  \multicolumn{1}{|c|}{Sampling rate:} \\
        \multicolumn{1}{|c|}{$N$} &  \multicolumn{1}{|c|}{$\Delta t$ ($\mu$s)} &  \multicolumn{1}{|c|}{$f_s$ (kHz)} \\
        \hline
        250 & 5.12 & 195 \\
        125 & 10.2 & 97.7 \\
        62 & 20.6 & 48.4 \\
        31 & 41.3 & 24.2 \\
        13 & 98.5 & 10.2 \\
        \hline
    \end{tabular}
    \end{center}
\end{table}

Then, we convert these power spectral densities into real time-domain current noise signals by performing an inverse discrete Fourier transform on each power spectral density (to which we applied randomly chosen phases, i.e. we add signals incoherently). Since we started in the frequency domain, we assume that the power spectral densities were obtained for signals measured with an record length $T_s$ and sampled with a sample interval $\Delta t$ (i.e. a sampling rate $f_s = 1/ \Delta t$).

Different sampling parameters are shown in Table \ref{tab:simuparam}. In reality, the record length $T_s$ would not be fixed but rather the number of samples would be fixed by the field-programmable gate arrays (FPGAs) that perform the analysis. We assume that with our effective area falling off at high energies that we won't be getting many single pulses that last longer than $T_s = 1.28$ ms, so when we are varying the sampling rate we are looking at the complexity of the FPGAs and the readout, i.e. a lower sampling rate for a lower complexity. In this work, we assume a sample interval of 5.12~$\mu$s, except mentioned otherwise.

\begin{figure}[ht]
\includegraphics[width=\columnwidth]{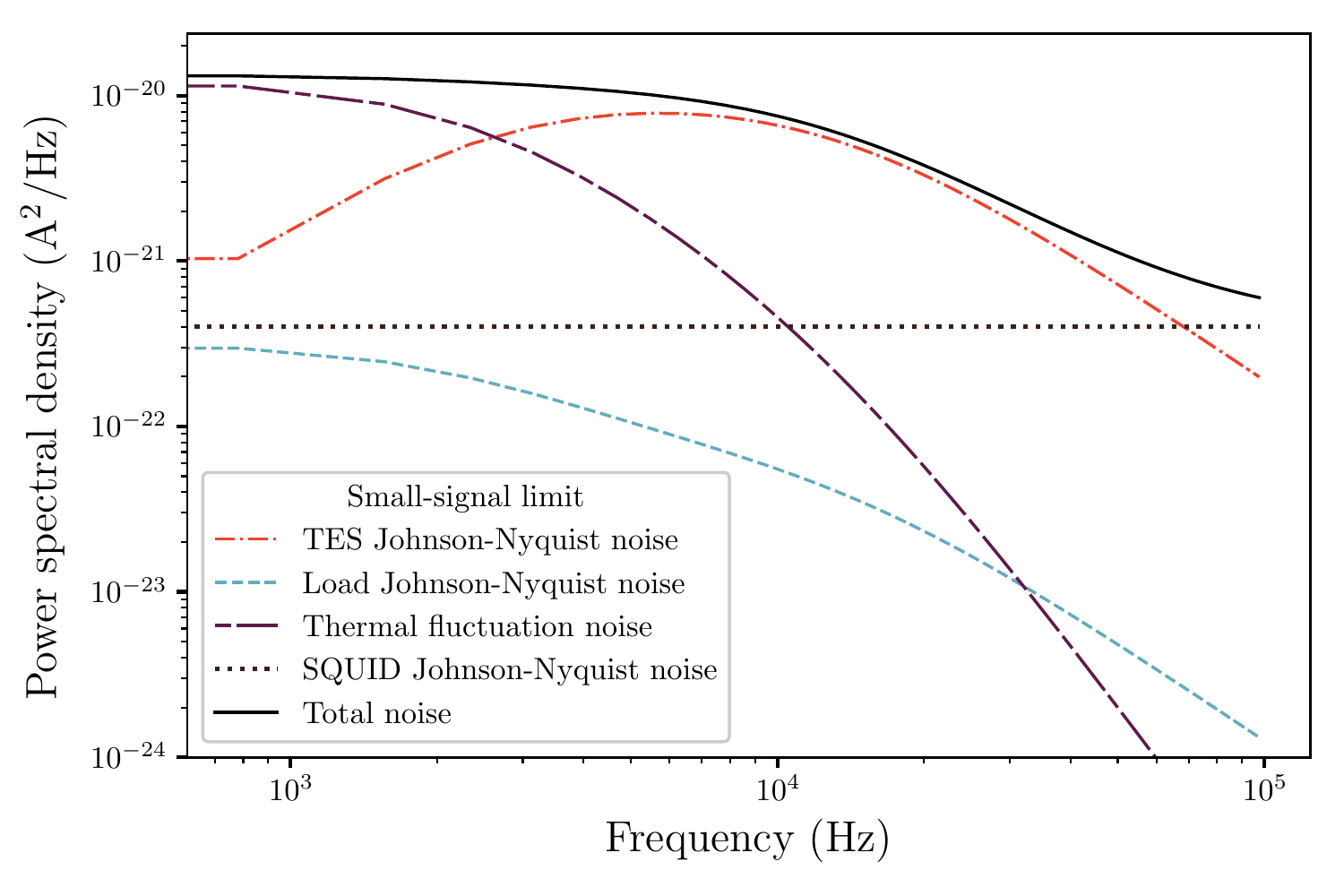}
\caption{Power spectral density of different sources of noise in a TES.}
\label{fig:current_noise_psd}
\end{figure}

\begin{figure}[ht]
\includegraphics[width=\columnwidth]{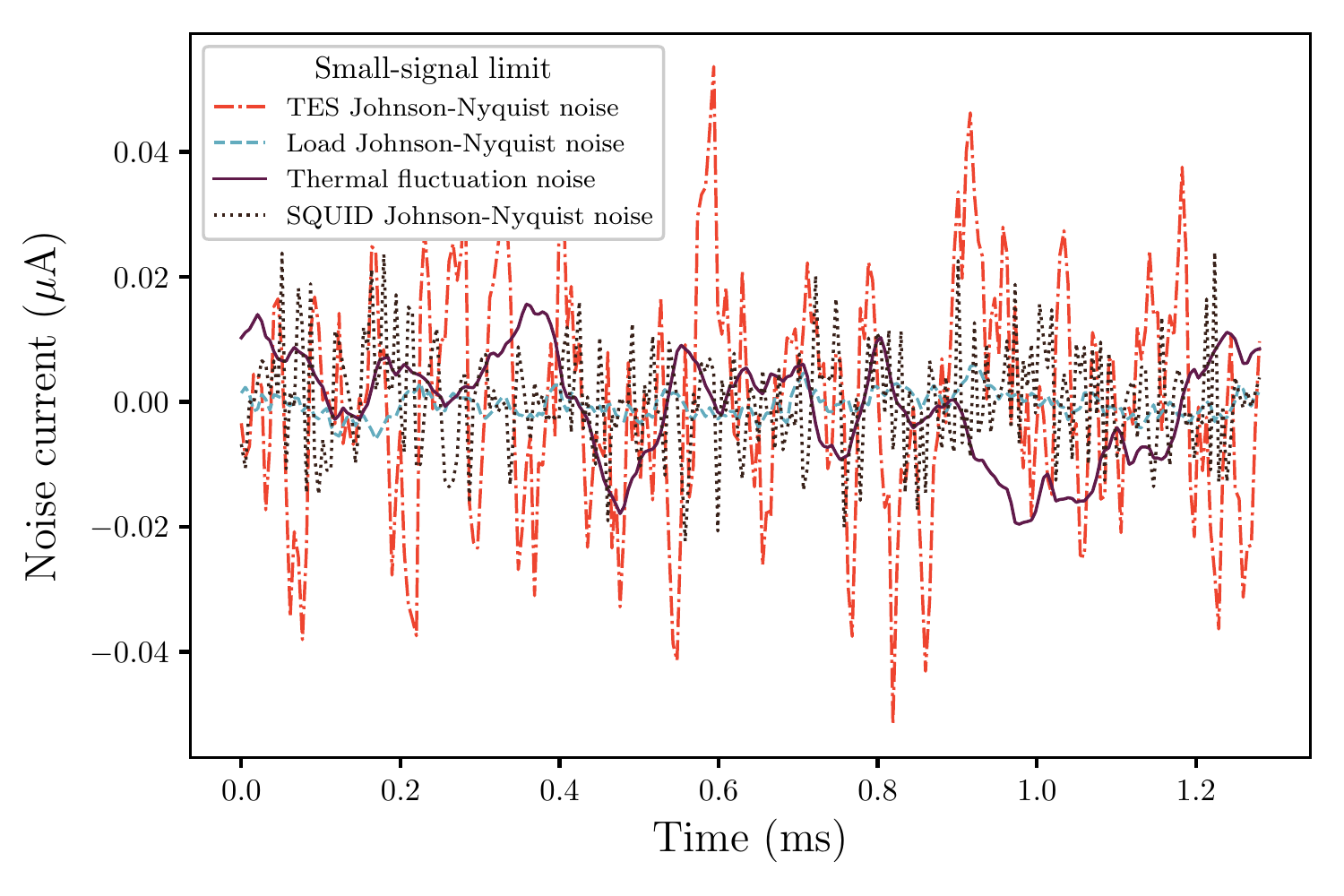}
\caption{Example of time-domain signals for different sources of current noise in a TES.}
\label{fig:current_noise_time_domain}
\end{figure}

In Fig.~\ref{fig:current_noise_psd}, we show the different power spectral densities calculated in Eq. (\ref{eq:PSD_JN_TES}) -- (\ref{eq:PSD_SQUID}). Since each time-domain noise signal is real, Fig.~\ref{fig:current_noise_psd} shows the positive-frequency part of each power spectrum (up to $f_s/2$), the latter being symmetric with respect to the zero frequency. One can see that the TES JN noise is suppressed at low frequencies because of electrothermal feedback and at high frequencies because of the circuit bandwidth. The roll-off of the TF noise at about $800$ Hz is due to the natural time constant $\tau$. In Fig. \ref{fig:current_noise_time_domain}, we plot each real time-domain noise signals.

\section{Measuring the energy and the arrival time of an incoming X-ray photon}

We now present our method to optimize energy and time resolution for photons hitting a TES.  We use the TES parameters outlined in Tab.~\ref{tab:physparam} with $L=48$ nH and a sample interval of 5.12~$\mu$s. We work with that inductance so that the onset of the pulse can be well resolved.

\subsection{High-energy photons}
As explained in Sec.~\ref{sec:simulEvents}, we propose to analyze current pulses differently, according to the photon energy. In this section, we develop a technique to measure energy and arrival time for high-energy photons (hard X-rays).

We first split a theoretical current pulse generated by a $7$-keV photon (see Sec.~\ref{sec:simulEvents}) in two parts, the ``onset'' (rising) part given by $I_{onset}(t)$, and the ``decay'' (falling) part given by $I_{decay}(t)$. The splitting is defined to be at $90 \%$ of the maximum pulse height; this enables to have enough points for curve fitting while avoiding saturation effects in the pulse. We do not expect the energy of the photon considered for the theoretical pulse to affect the resolution, as long as it has an energy large enough that the signal is saturating (because of the non-linear transition in a TES), i.e. $E_{\gamma} \gg E_{sat}$. Then, we build two model functions for curve-fitting as follows:
\begin{equation}
    I^\text{fit}_{onset}= k_{onset} I_{onset}(t-t_{onset}),
    \label{eq:yOnset}
\end{equation}
\begin{equation}
    I^\text{fit}_{decay}= I_{decay}(t-t_{decay}),
    \label{eq:yDecay}
\end{equation}
where $k_{onset}$, $t_{onset}$ and $t_{decay}$ are free parameters. 

\begin{figure}[ht]
\includegraphics[width=\columnwidth]{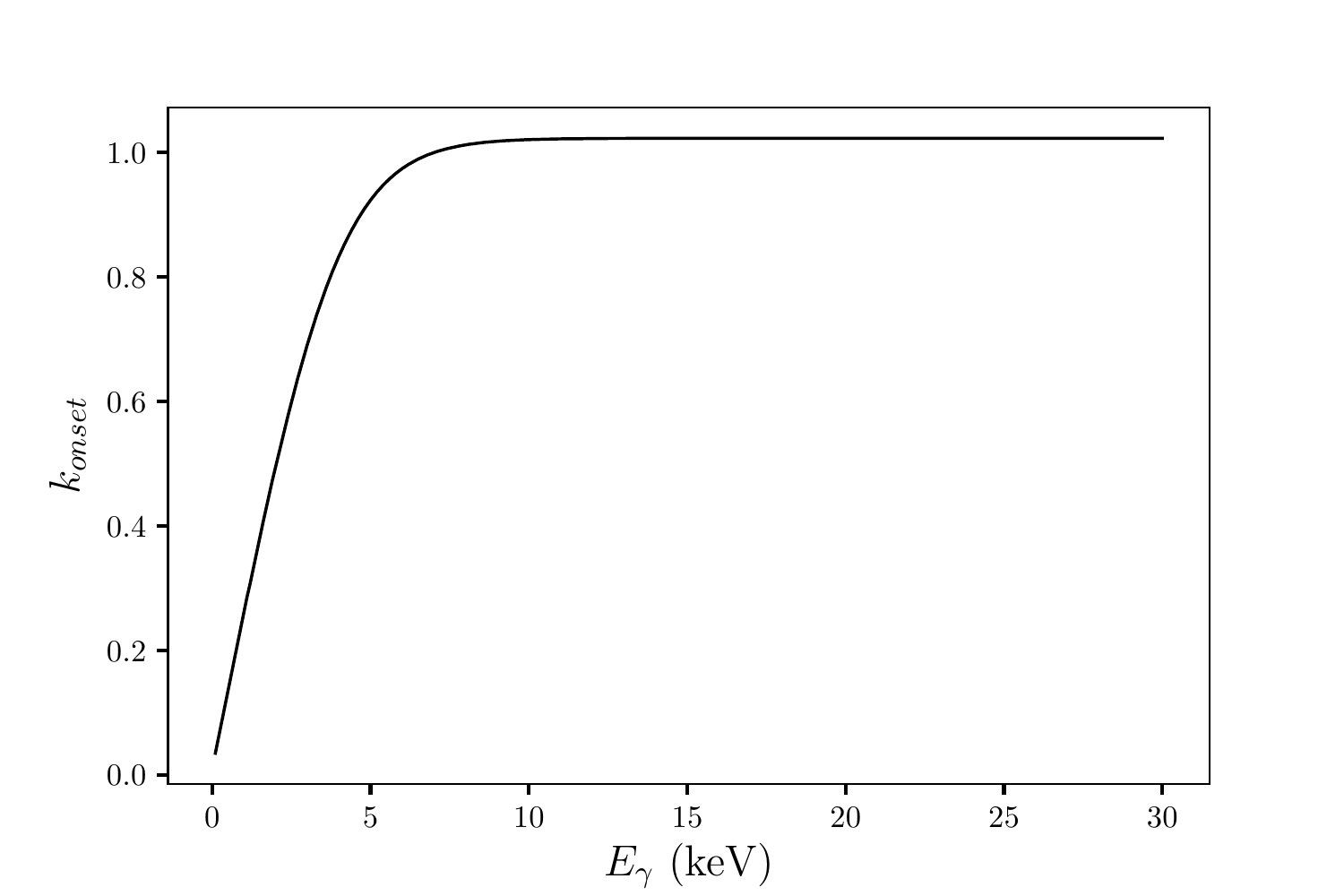}
\caption{$k_{onset}$ as a function of the energy of the incoming photon.}
\label{fig:k_onset}
\end{figure}

\begin{figure}[ht]
\includegraphics[width=\columnwidth]{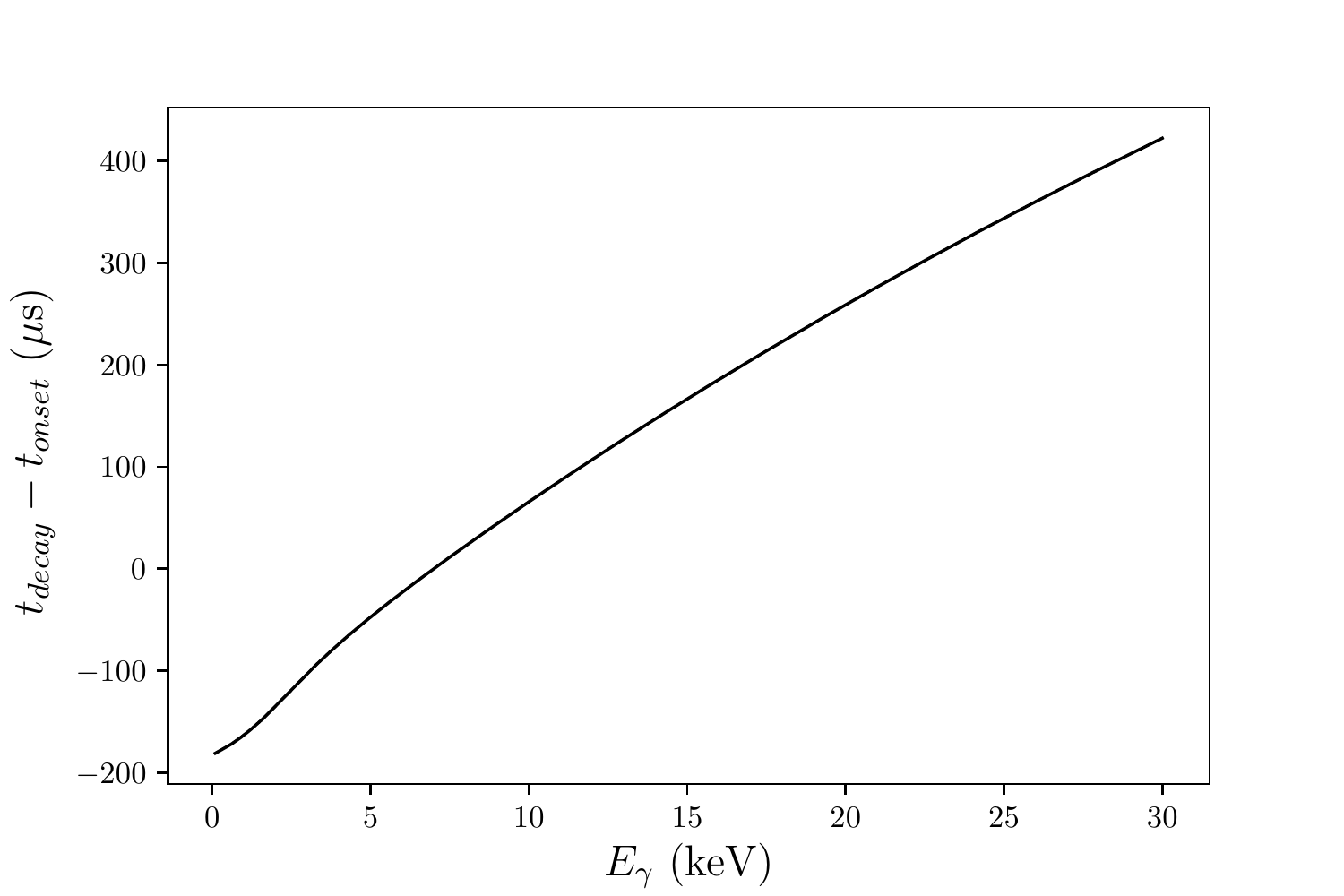}
\caption{$(t_{decay}-t_{onset})$ as a function of the energy of the incoming photon.}
\label{fig:t_decay_onset}
\end{figure}

We then simulate theoretical current pulses for photon energies between $0.1$ keV and $30$ keV, and for each energy we fit for $k_{onset}$, $t_{onset}$ and $t_{decay}$. We plot $k_{onset}$ and $(t_{decay}-t_{onset})$, respectively in Fig. \ref{fig:k_onset} and Fig. \ref{fig:t_decay_onset}.

The response of $k_{onset}$ is linear at low energies, which gives a first glimpse of the method used for low-energy photons. On the other hand, the response of $(t_{decay}-t_{onset})$ is linear at high energies, so this parameter is used to obtain the energy of the incoming photon. Therefore, we interpolate photon energies across $(t_{decay}-t_{onset})$, so we can retrieve the energy from the value of this parameter when analyzing simulated data. The parameter $t_{onset}$ is used to get the arrival time of the incoming photon. 

\begin{figure}[ht]
\includegraphics[width=\columnwidth]{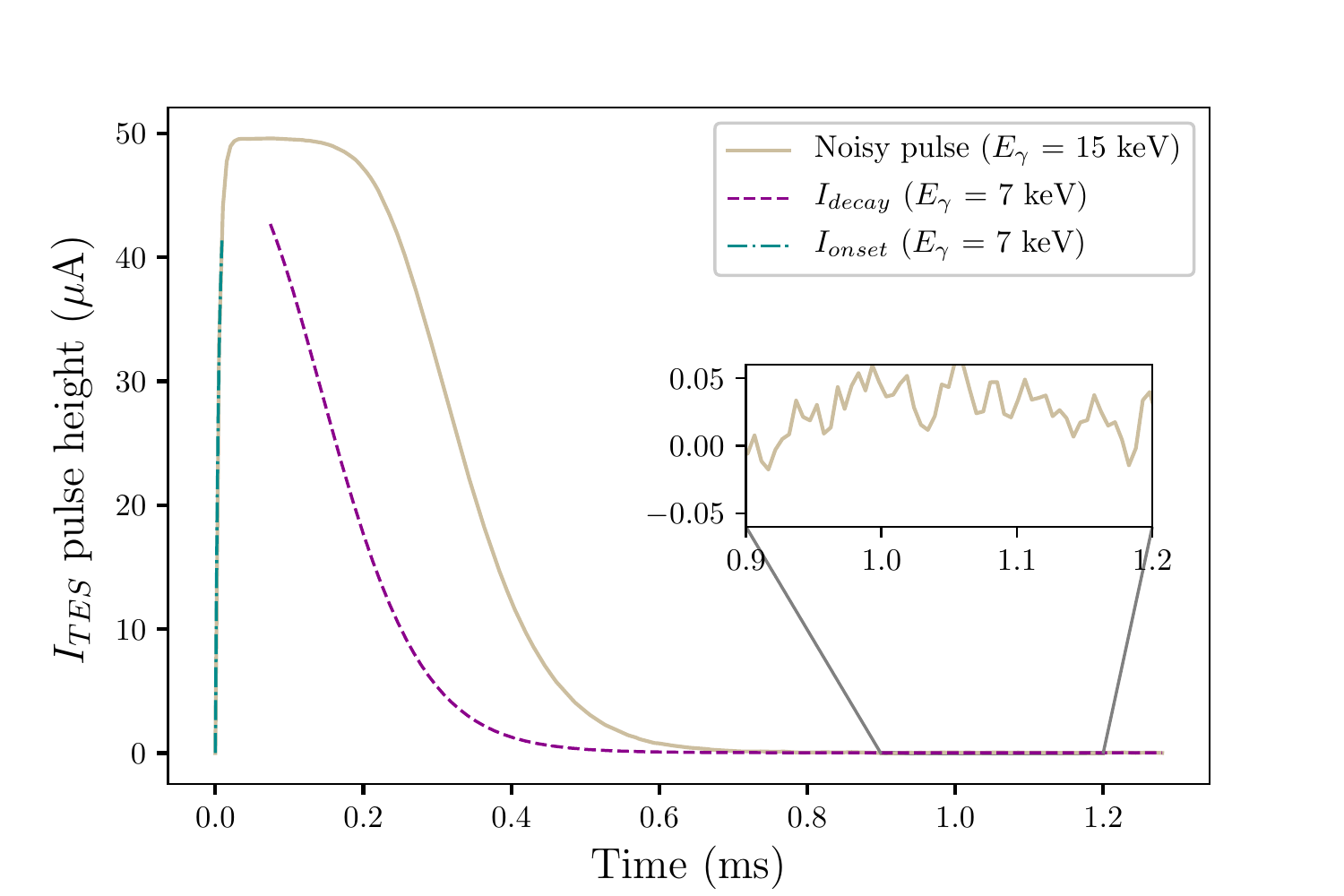}
\caption{Noisy current pulse (simulated data) for $E_\gamma = 15$ keV. A theoretical current pulse for $E_\gamma=7$ keV, splitted at $90~\%$ of its maximum height is also plotted for reference (cyan dash-dotted curve and magenta dashed curve). We zoom in (box) to attest the presence of noise in the simulation.}
\label{fig:he_photon_noisy_pulse}
\end{figure}

In order to predict the energy and time resolution expected when analyzing actual data, we run 1,000 simulations of sampled noisy current pulses, per photon energy. In these simulations, we do not assume perfect triggering, i.e. there is trigger-jitter that accounts for the random arrival phases of the photons -- photons do not arrive in perfect phase with the current sampling. We then get a distribution of values for $(t_{decay}-t_{onset})$ and $t_{onset}$, per energy. From these distributions, we calculate the full width at half maximum (FWHM) of the retrieved energy distribution, and the full duration at half maximum (FDHM) of the retrieved arrival-time distribution, per photon energy.  We also illustrate our method through an example in Fig.~\ref{fig:he_photon_noisy_pulse}. We simulate a sampled noisy current pulse for a $15$-keV photon; we obtain a FWHM of $2.37$ eV for the energy of the incoming photon, and a FDHM of $4.33$ ns for the arrival time of the incoming photon.

\subsection{Low-energy photons}
In this section, we develop a technique to measure energy and arrival time for low-energy photons (soft X-rays).

We first consider a theoretical current pulse for a $1$-keV photon, $I_{shape}(t)$ where the ``shape'' subscript refers to the fact that we fit the whole pulse shape in this method. From this theoretical pulse, we buid the curve-fitting model as follows:
\begin{equation}
    I^\text{fit}_{shape} = k_{shape} I_{shape}(t-t_{shape}),
    \label{eq:yShape}
\end{equation}
where $k_{shape}$, $t_{shape}$ are free parameters. 

\begin{figure}[ht]
\includegraphics[width=\columnwidth]{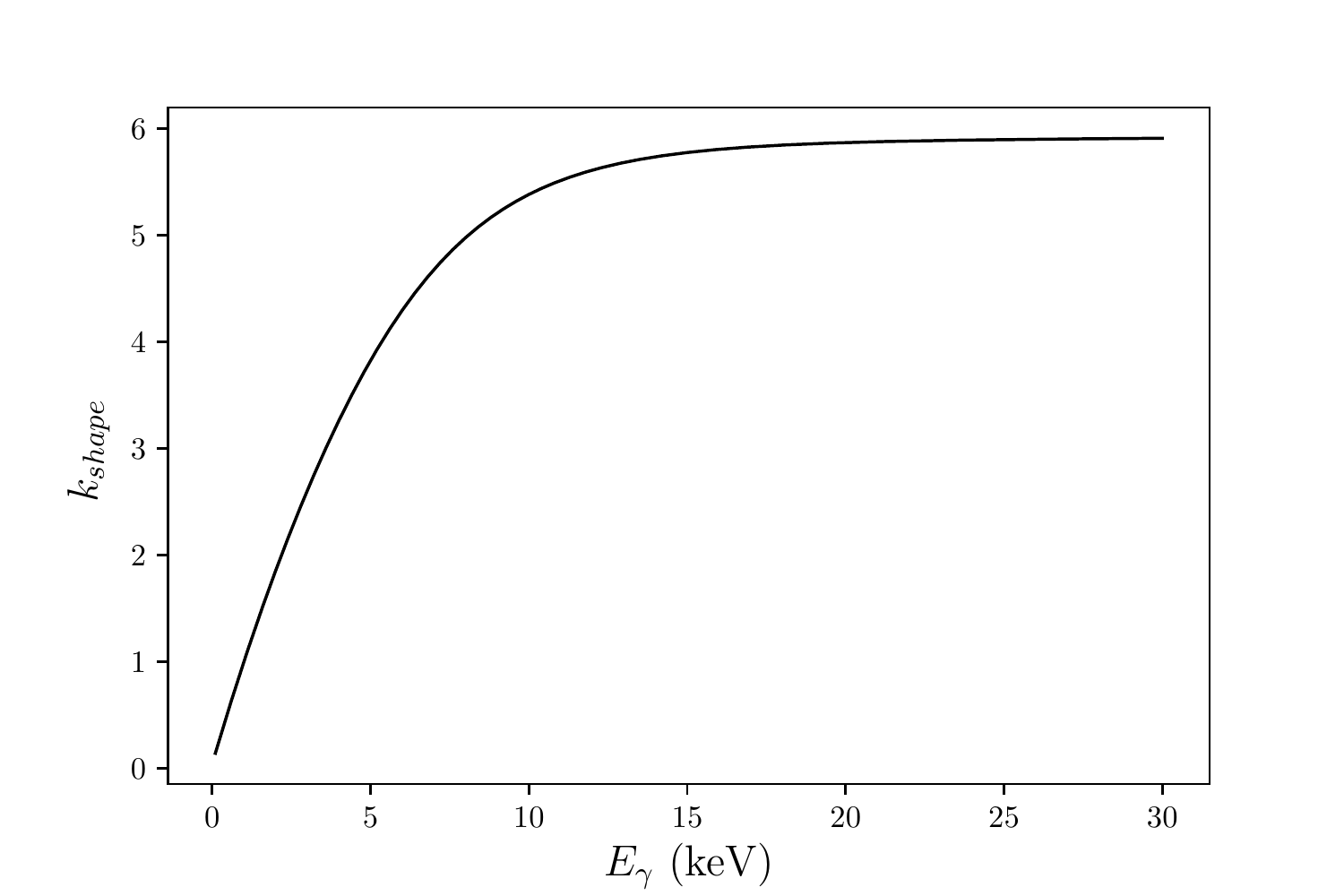}
\caption{$k_{shape}$ as a function of the energy of the incoming photon.}
\label{fig:k_shape}
\end{figure}

\begin{figure}[ht]
\includegraphics[width=\columnwidth]{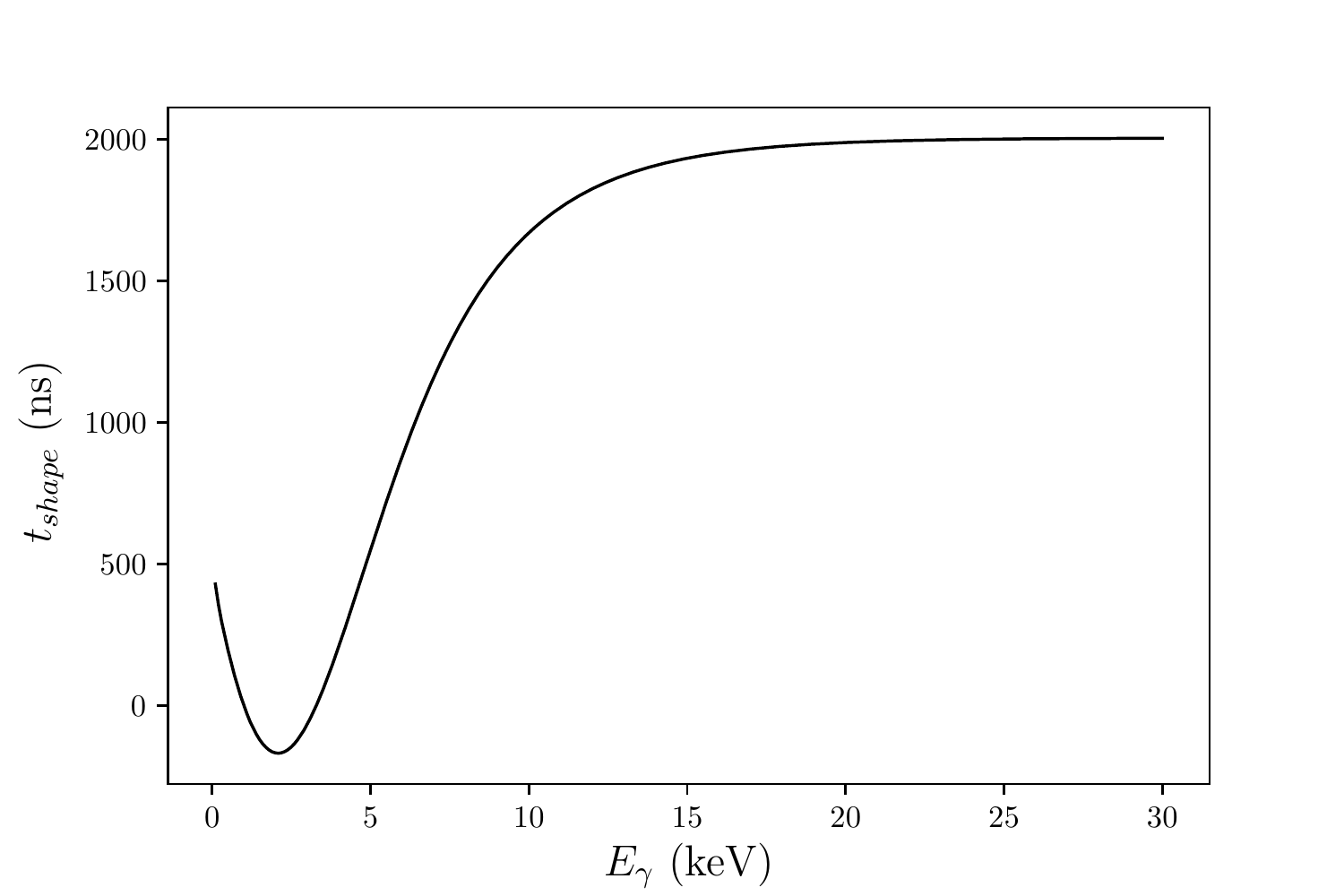}
\caption{$t_{shape}$ as a function of the energy of the incoming photon.}
\label{fig:t_shape}
\end{figure}

We then simulate theoretical current pulses for photon energies between $0.1$ keV and $30$ keV, and for each energy we fit for $k_{shape}$, $t_{shape}$.  We plot $k_{shape}$ and $t_{shape}$, respectively in Fig \ref{fig:k_shape}, Fig. \ref{fig:t_shape}.

The response of $k_{shape}$ is linear at low energies, so this parameter is used to obtain the energy of the incoming photon. Therefore, we interpolate photon energies across $k_{shape}$, so we can retrieve the energy from the value of this parameter when analyzing simulated data. The parameter $t_{shape}$ is used to get the arrival time of the incoming photon.

\begin{figure}[ht]
\includegraphics[width=\columnwidth]{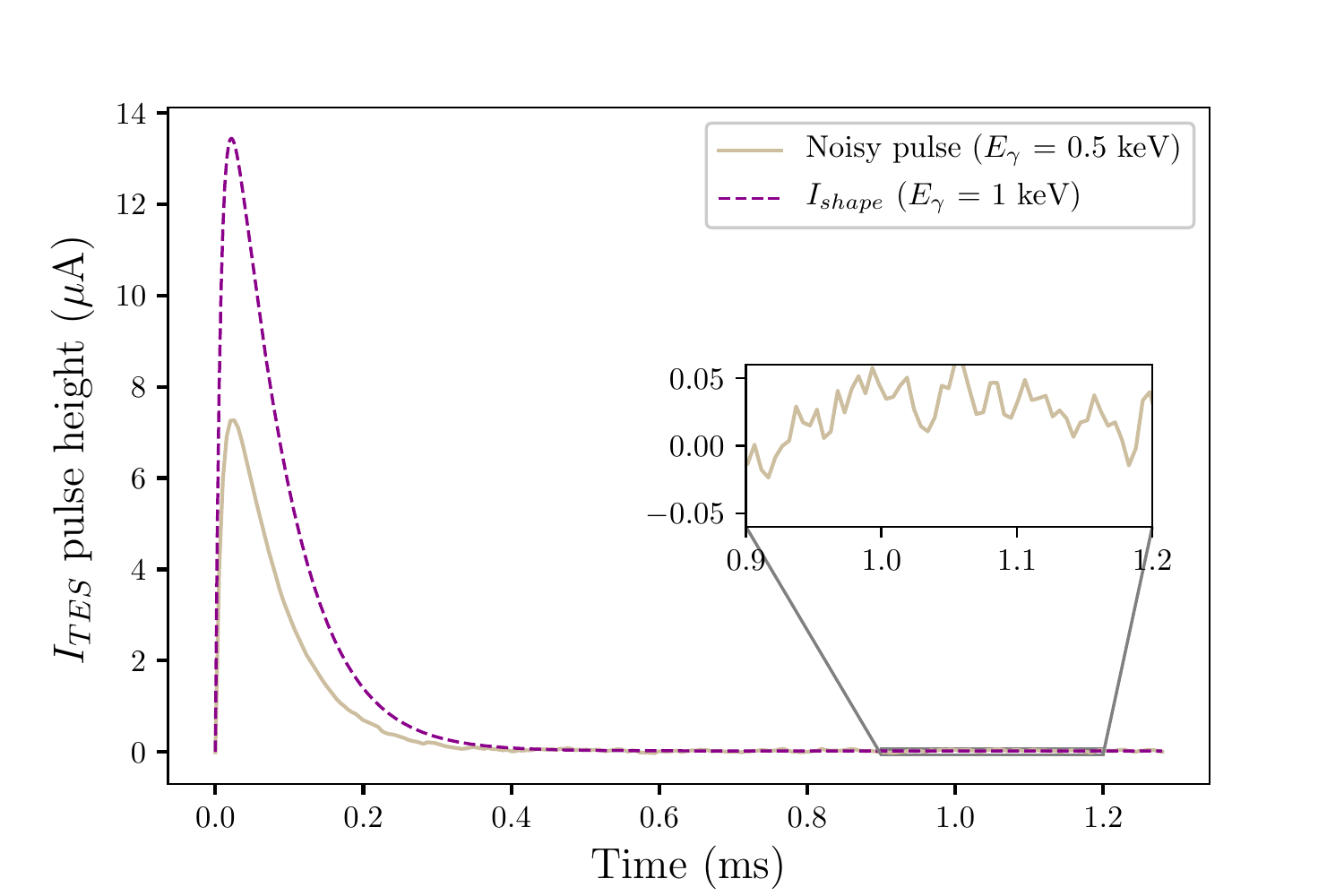}
\caption{Noisy current pulse (simulated data) for $E_\gamma = 0.5$ keV . A theoretical current pulse for $E_\gamma=1$ keV, splitted at $90~\%$ of its maximum height is also plotted for reference (magenta dashed curve). We zoom in (box) to attest the presence of noise in the simulation.}
\label{fig:le_photon_noisy_pulse}
\end{figure}

As before, in order to predict the energy and time resolution expected when analyzing actual data, we run 1,000 simulations of sampled noisy current pulses, per photon energy. In these simulations, we do not assume perfect triggering. We then get a distribution of values for $k_{shape}$ and $t_{shape}$, per energy. From these distributions, we calculate the FWHM of the retrieved energy distribution, and the FDHM of the retrieved arrival-time distribution, per photon energy.
We illustrate our method through an example in Fig.~\ref{fig:le_photon_noisy_pulse}. We simulate a sampled noisy current pulse for a $0.5$-keV photon; we obtain a FWHM of $1.39$ eV for the energy of the incoming photon, and a FDHM of $56$ ns for the arrival time of the incoming photon.

\subsection{Energy and arrival time resolution}
\begin{figure}[ht]
\includegraphics[width=\columnwidth]{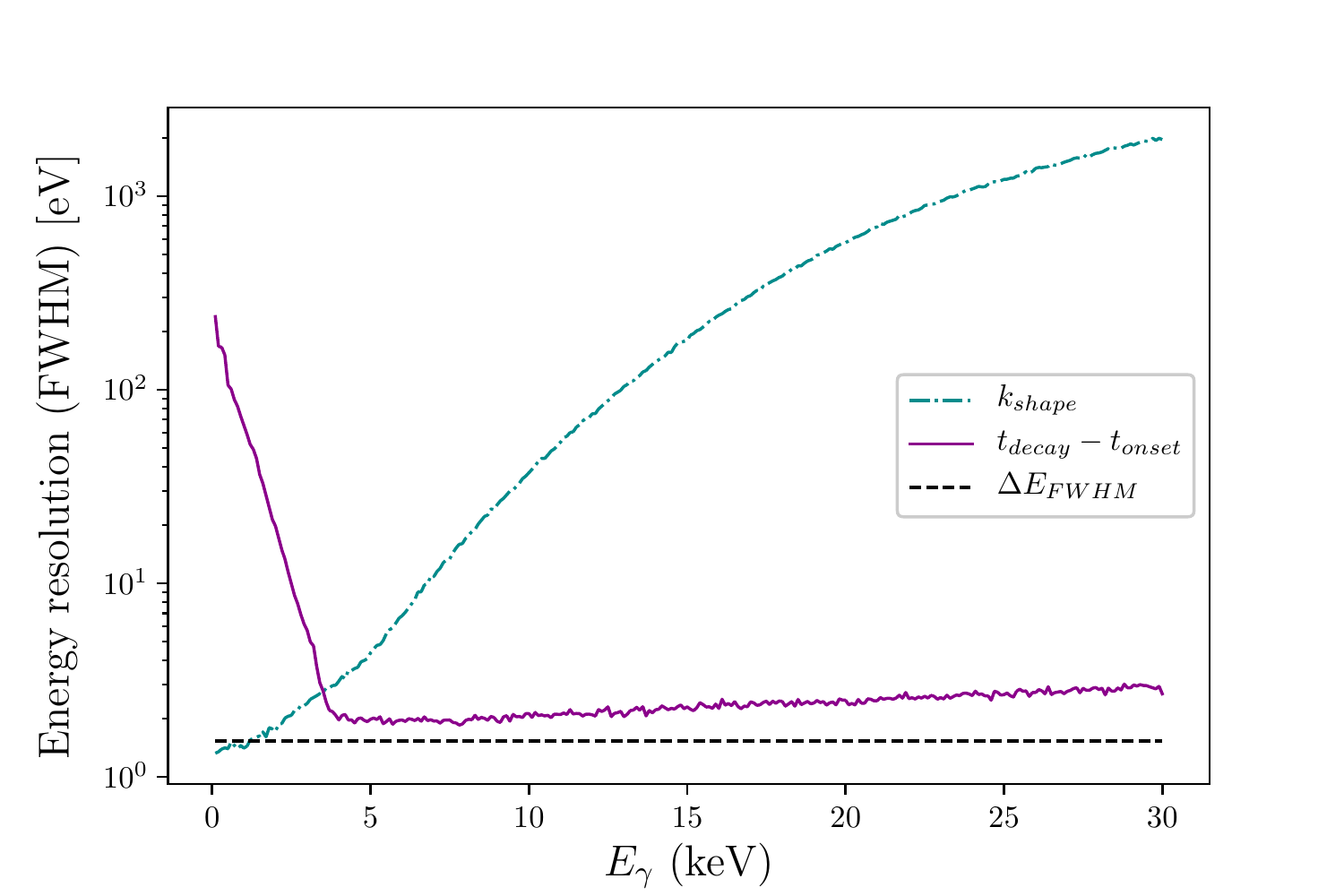}
\caption{Energy resolution (FWHM), derived from the parameters ($t_{decay}-t_{onset}$) and $k_{shape}$, as a function of the photon energy, $E_\gamma$.}
\label{fig:energy_resolution_250}
\end{figure}

We now summarize the energy resolutions for all photon energies in Fig.~\ref{fig:energy_resolution_250}. As one would expect, $k_{shape}$ gives the best energy resolution at low energies and $t_{decay}-t_{onset}$ gives it at high energies. We find the energy at which to switch methods to be $3.5$ keV. Our method enables us to reach an energy resolution (FWHM) between $1.32$ eV and $2.98$ eV, for energies between $0.1$ keV and $30$ keV. We also compare our results to the small-signal energy resolution (FWHM) of a transition-edge sensor, $\Delta E_{FWHM}$. In the important limit that the amplifier noise is negligible, $\Delta E_{FWHM}$ is given by Eq. (103) from Irwin (2005); for the TES considered in this work, it gives $\Delta E_{FWHM} = 1.53$ keV. One can notice in Fig.~\ref{fig:energy_resolution_250}, that for $E_{\gamma} < E_\text{sat}$ our method achieves a better resolution than $\Delta E_{FWHM}$. One needs to be cautious here since $\Delta E_{FWHM}$ is calculated from voltage noise instead of current noise. In the small-signal limit, noise is approximately stationary and we can interchange voltage noise and current noise (see section \ref{sec:noise}). However, in a TES, the noise is generally non-stationary, especially at large pulses. Therefore since we use stationary noise in this analysis, it is likely that our noise intensity is slightly underestimated, hence this resolution apparently better than the theoretical limit. As mentioned before, a full analysis of non-stationary noise is beyond the scope of this paper, but is necessary to improve the accuracy of our results. However, we expect our main conclusions to remain the same.

\begin{figure}[ht]
\includegraphics[width=\columnwidth]{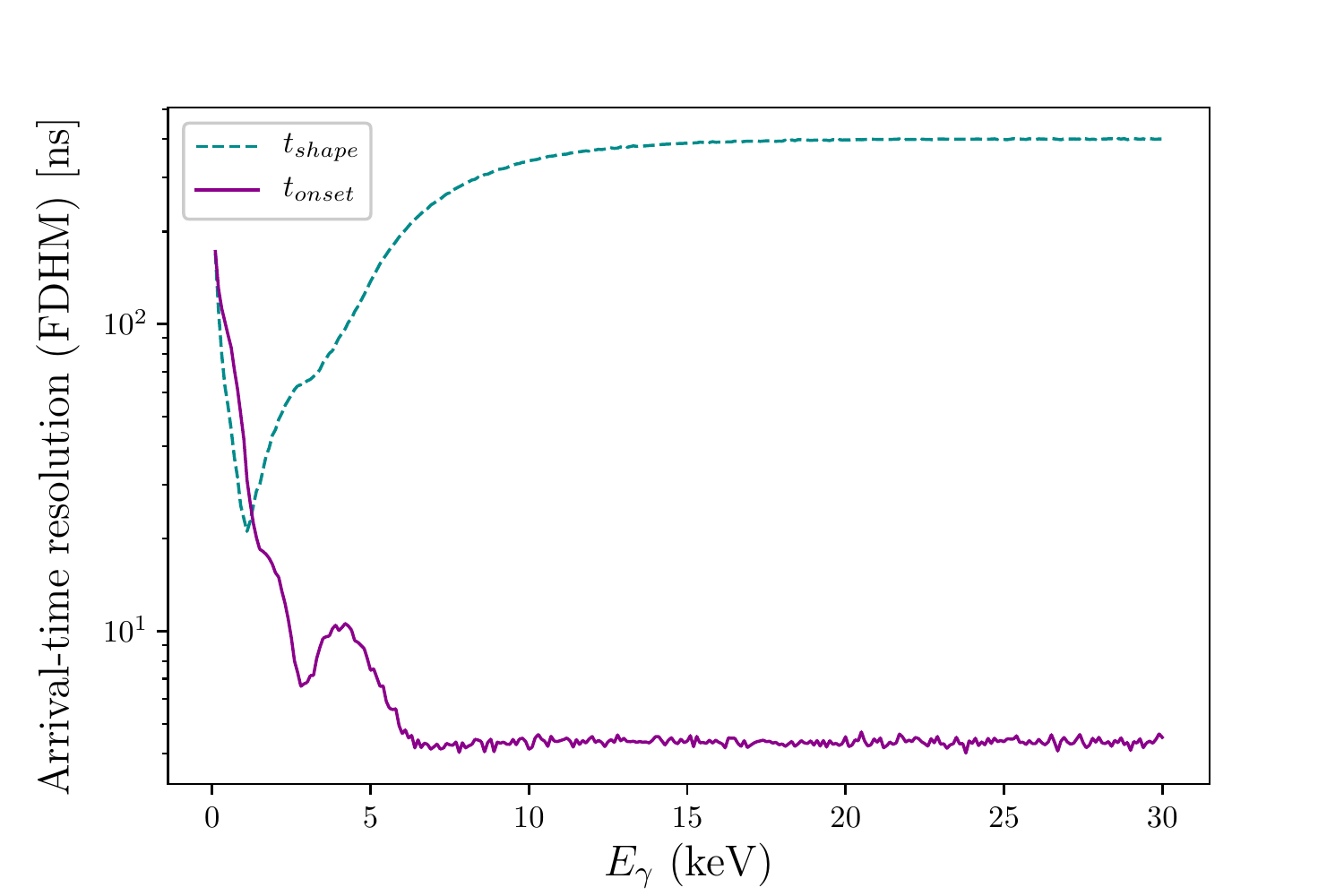}
\caption{Arrival-time resolution (FDHM), derived from the parameters $t_{onset}$ and $t_{shape}$, as a function of the photon energy, $E_\gamma$.}
\label{fig:time_resolution_250}
\end{figure}

Our method enables us to reach a time resolution (FDHM) between $163$ ns and $3.85$ ns, for energies between $0.1$ keV and $30$~keV (see Fig.~\ref{fig:time_resolution_250}). The energy at which the resolution from the two methods intersect is $1.3$ keV, which is roughly $E_\text{sat}$. This is expected, as the photon energy increases, the height of the current pulse increases and so does the signal-to-noise ratio. Since the signal is sampled with respect to time, the noise intensity has a greater effect on the measure of $t_{shape}$ than $k_{shape}$, the latter reflecting the pulse height.

These resolutions were obtained with a very simple method, which is promising for future analysis of X-ray-telescope data. However, pile-up events are known to alter the obtained resolution \cite{2014JLTP..176...16F}. Therefore, whenever a consistent photon energy cannot be retrieved with our method, the event should be treated separately as a pile-up event \cite{2012ITNS...59..366S}.

\section{Energy and arrival-time resolution for different parameters of a TES}

The energy and time resolution obtained depend on the parameters of the detector, such as the sampling rate, the heat capacity or the inductance. Consequently, we need to vary those parameters to see how resolution is impacted. Moreover, in reality the TES current signal will be digitized by the ADC. So we digitized our simulated signal at $195$ kHz with $16$ bits to see how the resolution would be affected; $16$ bits is assumed to be standard for spaceflight ADCs.

For simulations at $195$~kHz, we use the better energy resolution given by either ($t_{decay}-t_{onset}$) or $k_{shape}$, and the better time resolution given by either $t_{onset}$ or $t_{shape}$. However, for lower sampling rates or lower heat capacity, not enough points are present in the onset part for the curve fitting to obtain $t_{onset}$. Therefore, for those parameters, we use the better energy resolution given by either ($t_{decay}-t_{shape}$) or $k_{shape}$, and the time resolution is given by $t_{shape}$.

\subsection{Energy resolution}
\begin{figure}[ht]
\includegraphics[width=\columnwidth]{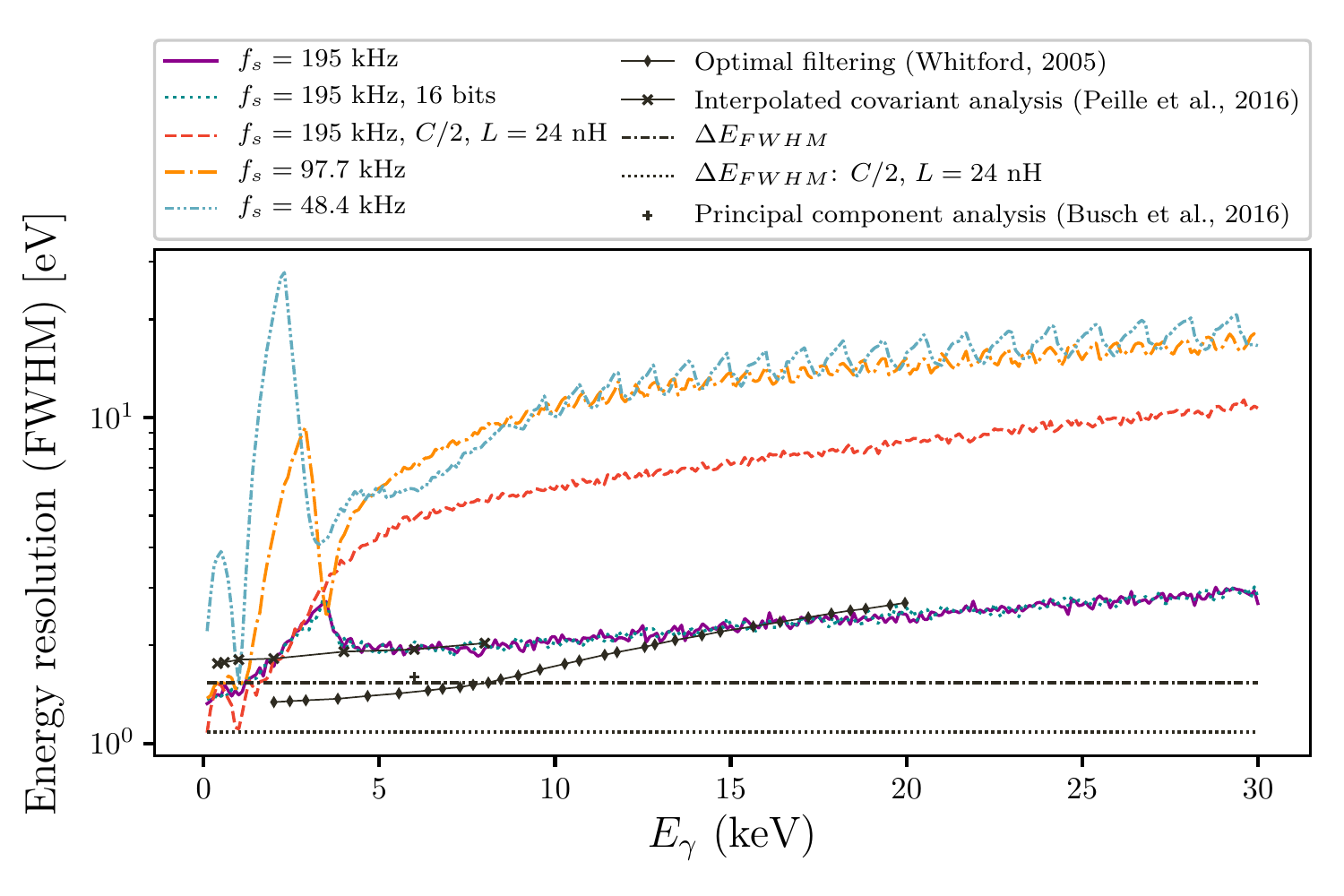}
\caption{Energy resolution (FWHM), for different TES parameters, as a function of photon energy. Results from this work are also compared to other techniques: optimal filtering \cite{2005NIMPA.555..255W}, principal component analysis \cite{2016JLTP..184..382B} and interpolated covariant analysis \cite{2016SPIE.9905E..5WP}}
\label{fig:energy_resolution}
\end{figure}

Energy resolutions for different TES parameters are showed in Fig.~\ref{fig:energy_resolution}. One can notice that digitizing the signal with $16$ bits has no significant effect on energy resolution and time resolution. This was expected since the digitization noise floor is much lower than the assumed TES noise.

Since $\Delta E_{FWHM}$ scales with $\sqrt{C}$ in the small-signal limit \cite{Irwin2005}, decreasing the heat capacity will improve the energy resolution at low energies. However our method still achieves a good energy resolution (below $10$ eV) at high energies, with half the heat capacity. In that case it is necessary to decrease the inductance to $L=24$~nH for the curve fitting. 

When the sampling rate is decreased further, trigger-jitter effects become more important, especially at low energies, and the energy resolution degrades. The wavy patterns of the energy resolution curve at high energy are artefacts due to the lower sampling rates, one can notice that the frequency of the pattern evolves with the sampling rate, as there are fewer and fewer points for curve fitting of the pulse. Yet, at $48.4$~kHz, our energy resolution is still below $20$~eV.

Finally, in Fig.~\ref{fig:energy_resolution}, we also compare our results to other common techniques used in TES analysis such as optimal filtering \cite{2005NIMPA.555..255W}, principal component analysis \cite{2016JLTP..184..382B} and interpolated covariant analysis \cite{2016SPIE.9905E..5WP}. Those techniques perform better at lower energies than our method. However, they also perform better than the theoretical FWHM (which includes most noise sources), this is most likely due to the fact that TES parameters used in those studies are different from the ones in ours. In Whitford (2005), it is also not clear which types of noise are considered in the analysis and if trigger-jitter (and sampling) effects are taken into account. Another pertinent comparative study of energy resolution retrieved with different techniques \cite{2016SPIE.9905E..5WP}, for the X-ray Integral Field Unit microcalorimeter on board the Athena X-ray observatory, shows energy resolutions between [$1.76$ -- $1.82$] eV and [$2.03$ -- $2.18$] eV, respectively for energies ranging between $0.2$ keV and $8$ keV. However, to our current knowledge, there doesn't exist complete TES-energy-resolution studies over an energy range as broad as ours. Therefore, it is not easy to accurately compare our results to other methods. Nevertheless, it seems reasonable to argue that our method gives similar results, especially at low energies, with previous methods but with possible improvements in terms of energy resolution (especially at higher energies) and computational cost.

\subsection{Arrival-time resolution}
Changing the TES parameters affects the time resolution as well (especially since at lower sampling rates or heat capacity only $t_{shape}$ can be computed), but in any case it remains far below $1~\mu$s. We compare our results to the time resolution achived by code-division multiplexing \cite{2016ApPhL.109k2604M}. Our results perform about $70$ times better, this could be due to the difference in the assumed TES parameters and noise. 

As for the energy, to our current knowledge, there does not exist complete TES-time-resolution studies over an energy range as broad as ours. Therefore, it is not easy to accurately compare our results to other methods. The achieved time resolutions with our method are very promising for X-ray-telescope applications.

\begin{figure}[ht]
\includegraphics[width=\columnwidth]{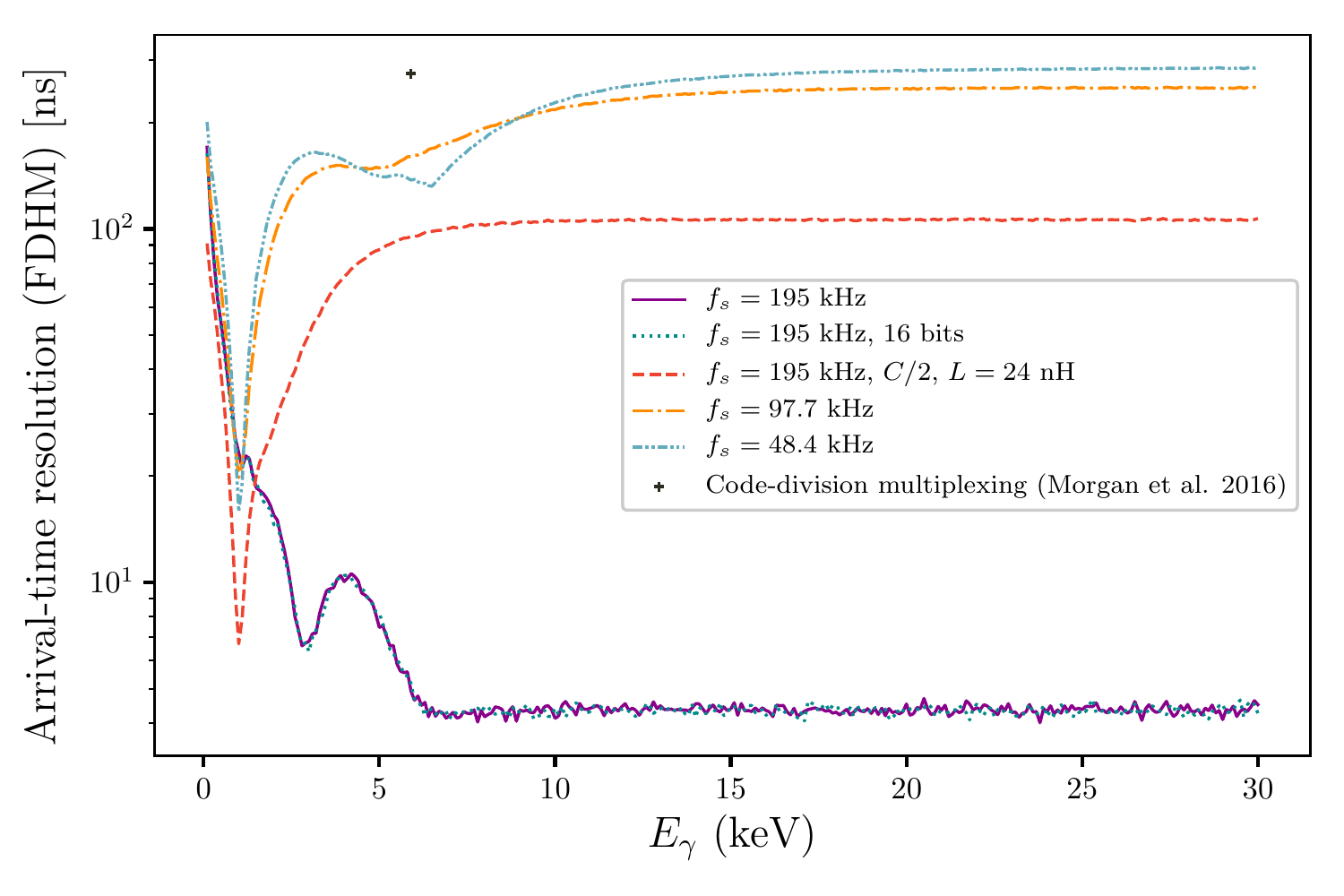}
\caption{Arrival-time resolution (FDHM), for different TES parameters, as a function of photon energy. Results from this work are also compared to a result from code-division multiplexing \cite{2016ApPhL.109k2604M}}
\label{fig:time_resolution}
\end{figure}

\section{Conclusion}
TESs are key devices for future X-ray telescopes \cite{2017ITAS...2749839D} and for astrophysics research in general. Retrieving the energy and the arrival time of an X-ray photon detected by a TES has been a research topic of great interest for about 25 years \cite{1995PhDT.........2I}. However, most techniques developed for that purpose have a high computational cost (see Table 2 in \cite{2016SPIE.9905E..5WP} for example), which is problematic for space applications where limited computational power will be available. We have successfully developed a method that enables to optimize the measurements of the energy and the arrival time of photons detected by a TES, using a simple pulse analysis. Such a method is promising for future X-ray data processing as it should have a lighter computational cost than other common techniques.

As a future work, to improve the accuracy of our results, especially at high energies, it will be necessary to include a thorough description of non-stationary noise in our simulations. Furthermore, at high count rates, the individual incoming photons will have closer and closer arrival times. This leads to pile-up effects that will affect the reconstruction of pulse and so the energy and time resolution \cite{2009AIPC.1185..450D}. Consequently, it will also be needed to develop a robust way to identify pile-up events -- this should be performed by the on-board central processing unit (CPU) rather than by the FPGA. Last but not least, we will determine how to efficiently deal with real-time processing, and therefore enable its implementation in future X-ray telescopes.

\subsection*{Disclosures}
The authors have no relevant financial interests in this article nor any other potential conflicts of interest.

\acknowledgments 
This work was supported by the Natural Sciences and Engineering Research Council of Canada, the Canada Foundation for Innovation, and the British Columbia Knowledge Development Fund.

We thank the anonymous referee for thoroughly reading the manuscript and providing very valuable comments that helped to shape this research. Particularly, we thank the referee for providing helpful insights regarding TES noise and trigger-jitter effects.

Simulations were run on: UBC Advanced Research Computing, ``UBC ARC Sockeye.'' UBC Advanced Research Computing, 2019, \url{https://doi.org/10.14288/SOCKEYE}. Python 3 \cite{10.5555/1593511}, Jupyter Notebooks \cite{soton403913}, NumPy \cite{2020NumPy-Array}, SciPy \cite{2020SciPy-NMeth}, and Matplotlib \cite{4160265} were used in this work. This research has made use of National Aeronautics and Space Administration’s Astrophysics Data System Bibliographic Services.

The archived version of the code can be freely accessed and executed through Code Ocean.


\bibliography{bibliography}   
\bibliographystyle{spiejour}   


\vspace{2ex}\noindent\textbf{Paul Ripoche} is a PhD student at the University of British Columbia. His current research interests include compact objects, carbon stars, and instrumentation for X-ray astronomy.

\vspace{2ex}\noindent\textbf{Jeremy Heyl} is a professor at the University of British Columbia.  His current research interests include black holes, neutron stars, white dwarfs, stellar populations and instrumentation for X-ray astronomy

\listoffigures
\listoftables

\end{spacing}
\end{document}